\documentclass[11pt]{article}
\usepackage{latexsym}
\usepackage{psfig}
\usepackage[latin2]{inputenc}
\font\pici=cmssi8

\newcommand{\qed}{$\Box$}
\newtheorem{theorem}{Theorem}
\newtheorem{lemma}[theorem]{Lemma}
\newtheorem{definition}[theorem]{Definition}
\newtheorem{corollary}[theorem]{Corollary}

\newtheorem{example}[theorem]{Example}

\newenvironment{proof}
     {\pagebreak[1]{\narrower\noindent {\bf
Proof:\quad\nopagebreak}}}{\qed}

\title{Defying Dimensions Modulo 6\\
{\tiny Preliminary Version}}
\author{Vince Grolmusz \thanks{Department of Computer Science, 
E\"otv\"os University, Budapest,
P\'azm\'any P. stny. 1/C, H-1117 Budapest, Hungary; E-mail:
grolmusz@cs.elte.hu} }

\begin{document}

\date{}

\maketitle

\begin{abstract} 
We show that a certain representation of the matrix-product can be computed with $n^{o(1)}$ multiplications. We also show, that siumilar representations of matrices can be compressed enormously.

\end{abstract}

\section{Introduction}

Gigantic matrices emerge in applications in numerous fields of mathematics, computer science, statistics and engineering. The storage and the basic linear-algebraic operations of these matrices is a difficult task which should be handled in data-mining, signal-processing, image-processing and various other applications day bay day. 

Here we describe a method which, by the simplest tools, may help in these tasks. We describe how to store large matrices or do linear algebraic operations on them if it is sufficient to retrieve the matrices or the results of the linear-algebraic operations only in a certain form, called 1-a-strong representation modulo 6 in \cite{G-mtxprod}. 

For $n\times n$ matrices $A$ and $B$ with elements from set $\{0,1,2,3,4,5\}$, our results include:
\begin{itemize}
\item The computation of $n^{o(1)}\times n^{o(1)}$ matrix $A'$ with elements from set $\{0,1,2,3,4,5\}$, such that from $A'$ one can retrieve the 1-a-strong representation of the $n\times n$ matrix $A$; here both operations are simple linear transformations (here $o(1)$ denotes a sequence of positive numbers going to 0 while $n$ goes to the infity).

\item The computation of the 1-a-strong representation of the matrix-product $AB$ with $n^{o(1)}$ multiplications, significantly improving our earlier result of computing the 1-a-strong representation of the matrix-product with $n^{2+o(1)}$ multiplications \cite{G-mtxprod}.

\end{itemize}

\subsection{Earlier Results: Matrix Product}

The matrix multiplication is a basic operation in mathematics in applications in almost every branch of mathematics itself, and also in the science and engineering in general. An important problem is finding algorithms for fast matrix multiplication. The natural algorithm for computing the product of two $n\times n$ matrices uses $n^3$ multiplications. The first, surprising algorithm for fast matrix multiplication was the recursive method of Strassen \cite{Strassen}, with $O(n^{2.81})$ multiplications. After a long line of results, the best known algorithm today was given by Coppersmith and Winograd \cite{Coppersmith-Winograd}, requiring only $O(n^{2.376})$ multiplications. Some of these methods can be applied successfully in practice for the multiplication of large matrices \cite{Bailey}.

The best lower bounds for the number of needed multiplications are between $2.5n^2$ and $3n^2$, depending on the underlying fields (see \cite{Blaser}, \cite{Bshouty}, \cite{Shpilka01}). A celebrated result of Raz \cite{razlower} is an $\Omega(n^2\log n)$ lower bound for the number of multiplications, if only bounded scalar multipliers can be used in the algorithm.

In \cite{G-mtxprod} we gave an algorithm with $n^{2+o(1)}$ multiplications for computing a {\em representation} of the matrix product modulo non-prime power composite numbers (e.g., 6). The algorithm was a straightforward application of a method of computing the representation of the dot-product of two length-$n$ vectors with only $n^{o(1)}$ multiplications.

In the present work, we significantly improve the results of \cite{G-mtxprod}, we gave an algorithm for computing a representation of the product of two $n\times n$ matrices with only $n^{o(1)}$ multiplications.

\subsubsection{Why do we count only the multiplications?}

In algebraic algorithms it is quite usual to count only the multiplications in a computation. The reason for this is that the multiplication is {\em considered} to be a harder operation than the addition in most practical applications, and moreover, the multiplication is {\em proven} to be harder in most theoretical models of computation. 

For example, computing the PARITY is reduced to computing the multiplication of two $n$-bit sequences, and, consequently, two $n$-bit sequences cannot be multiplied on a polynomial-size, constant-depth Boolean circuit \cite{FSS}, while it is well known, that two $n$-bit sequences can be added in such a circuit. 

Another example is the communication complexity \cite{NiKu} of the computation of multi-variable polynomials. Here two players, Alice and Bob, want to co-operatively compute the value of a $2n$-variable polynomial $f(x,y)$ modulo $m$, where the polynomial $f$ is known for both players, while $x$ is known only for Alice, and $y$ is known only for Bob. If $f$ is a linear polynomial, then they can compute $f(x,y)$ with communicating only $O(\log m)$ bits: Indeed, if 
$$f(x,y)=\sum_i a_ix_i+\sum_jb_jy_j,$$
then Alice communicates to Bob the value $$\sum_i a_ix_i\bmod{m},$$
then Bob will know the value of $f(x,y)$. Similarly, if $f(x,y)$ can be given as the sum of $u$ products, each with $v$ clauses:
$$f(x,y)=\sum_{\mu=1}^u\prod_{\nu=1}^v\sum_{i,j}a_{i\mu\nu}x_i+b_{j\mu\nu}y_j,$$
then $f(x,y)$ can be computed with $O(uv\log m)$ bits of communication. This example also shows that in communication complexity the multiplications may be harder than the additions. Chor and Goldreich \cite{CG} proved that even approximating the dot-product function
$$f(x,y)=\sum_{i=1}^nx_iy_i$$
needs $\Omega(n)$ bits of communication; consequently, multiplications are really hard in the communication model.

\subsection{Earlier Results: Matrix Storage}

There are lots of results concerning sparse matrix-storage and operations. Here we do not assume anything about the sparsity of our input matrices.

We should mention here the works of Frieze and Kannan \cite{Frieze-Kannan} and \cite{Frieze-Kannan-Vempala} on approximations of large matrices with small rank matrices with very fast algorithms.

\section{Preliminaries}

First we need to review several definitions and statements from \cite{G-mtxprod}.

\subsection{A-strong representations}

In \cite{G-six} we gave the definition of the {\em a-strong (i.e., alternative-strong) representation}  of polynomials. Here we define the {\em alternative}, and the {\em 0-a-strong} and the {\em 1-a-strong} representations of polynomials. Note that the 0-a-strong representation, defined here, coincides with the a-strong representation of the paper \cite{G-six}.

Note also, that for prime or prime-power moduli, polynomials and their
representations (defined below), coincide. Perhaps that is the reason that such representations were not defined before.

\begin{definition}\label{a-strong}
Let $m$ be a composite number $m=p_1^{e_1}p_2^{e_2}\cdots 
p_\ell^{e_\ell}$. Let $Z_m$ denote the ring of modulo $m$ integers. Let 
$f$ be a polynomial of $n$ variables over $Z_m$:
$$f(x_1,x_2,\ldots,x_n)=\sum_{I\subset \{1,2,\ldots,n\}}a_Ix_I,$$
where $a_I\in Z_m$, $x_I=\prod_{i\in I}x_i$. Then we say that
$$g(x_1,x_2,\ldots,x_n)=\sum_{I\subset \{1,2,\ldots,n\}}b_Ix_I,$$
is an
\begin{itemize}
\item {\em alternative representation} of $f$ modulo $m$, if
$$\forall I\subset\{1,2,\ldots,n\}\ \  \exists j\in \{1,2,\ldots,\ell\}: 
\ \ a_I\equiv b_I \pmod{p_j^{e_j}};$$
\item {\em 0-a-strong representation} of $f$ modulo $m$, if
it is an alternative representation, and, furthermore,
 if for some $i$, $a_I\not\equiv b_I \pmod{p_i^{e_i}},$ then 
$b_I\equiv 0\pmod{p_i^{e_i}};$
\item {\em 1-a-strong representation} of $f$ modulo $m$, if
it is an alternative representation, and, furthermore,
 if for some $i$, $a_I\not\equiv b_I \pmod{p_i^{e_i}},$ then 
$a_I\equiv 0\pmod{m};$

\end{itemize}
\end{definition}

\begin{example}Let $m=6$, and let $f(x_1,x_2,x_3)=x_1x_2+x_2x_3+x_1x_3$,
then 
$$g(x_1,x_2,,x_3)=3x_1x_2+4x_2x_3+x_1x_3$$ is a 0-a-strong representation 
of $f$ modulo 6;
$$g(x_1,x_2,,x_3)=x_1x_2+x_2x_3+x_1x_3+3x_1^2+4x_2$$
is a 1-a-strong representation of $f$ modulo 6;
$$g(x_1,x_2,,x_3)=3x_1x_2+4x_2x_3+x_1x_3+3x_1^2+4x_2$$
is an alternative representation modulo 6.
\end{example}

In other words, for modulus 6, in the alternative representation, each coefficient is correct either modulo 2 or modulo 3, but not necessarily both. 

In the 0-a-strong representation, the 0 coefficients are always correct both modulo 2 and 3, the non-zeroes are allowed to be correct either modulo 2 or 3, and if they are not correct modulo one of them, say 2, then they should be 0 mod 2.

In the 1-a-strong representation, the non-zero coefficients of $f$ are correct for both moduli in $g$, but the zero coefficients of $f$ can be non-zero either modulo 2 or modulo 3 in $g$, but not both.

\begin{example}
Let $m=6$. Then $0=xy-3xy+2xy$ is {\bf not} a 1-a-strong representation of $xy$. Similarly, polynomial $f+2g+3h$ is a mod 6 1-a-strong representation of polynomial $f$ if and only if $g$ and $h$ do not have common monomials with $f$, and $g$ does not have common monomials with $h$.
\end{example}

\subsection{Previous results for a-strong representations}

We considered elementary symmetric polynomials 
$$S_n^k=\sum_{{I\subset \{1,2,\ldots,n\}}\atop{|I|=k}}\prod_{i\in 
I}x_i$$ in \cite{G-six}, and proved that for constant $k$'s, 0-a-strong
representations of elementary symmetric polynomials $S_n^k$ can be
computed dramatically faster over non-prime-power composites than over
primes: we gave a depth-3 multilinear arithmetic circuit with
sub-polynomial number of multiplications (i.e., $n^\varepsilon, \forall
\varepsilon>0$), while over fields or prime moduli computing these polynomials on depth-3 multilinear circuits 
needs polynomial (i.e., $n^{\Omega(1)}$) multiplications.

Here depth-3 multi-linear, homogeneous arithmetic circuits computes polynomials of the form $$\sum_{j=1}^t\prod_{k=1}^\ell\sum_{i=1}^na_{jki}x_i\eqno{(1)}.$$  These circuits are sometimes called $\Sigma\Pi\Sigma$ circuits.

In \cite{G-six}, we proved the following theorem:

\begin{theorem}[\cite{G-six}]\label{fo-1}
\begin{itemize}
\item[(i)] Let $m=p_1p_2$, where $p_1\ne p_2$ are primes. Then a degree-2 0-a-strong representation of 
$$S_n^2(x,y)=\sum_{i,j\in\{1,2,\ldots,n\}\atop{i\ne j}}x_iy_j,$$
 modulo $m$:
$$\sum_{i,j\in\{1,2,\ldots,n\}\atop{i\ne j}}a_{ij}x_iy_j\eqno{(2)}$$
 can be computed on a bilinear $\Sigma\Pi\Sigma$ circuit of size
$$\exp(O(\sqrt{\log n\log \log n})).$$
Moreover, this representation satisfies that $\forall i\ne j: a_{ij}=a_{ji}$.

\item[(ii)] Let the prime decomposition of $m=p_1^{e_1}p_2^{e_2}\cdots p_r^{e_r}$. Then a degree-2 0-a-strong representation of $S_n^2(x,y)$ modulo $m$ of the form (1) can be computed on a bilinear $\Sigma\Pi\Sigma$ circuit of size
$$\exp\bigg(O\bigg({\sqrt[r]{\log n(\log \log n)^{r-1}}}\bigg)\bigg).$$
Moreover, this representation satisfies that $\forall i\ne j: a_{ij}=a_{ji}$.

\end{itemize}
\end{theorem}\qed

\begin{corollary}\label{egy}
The 0-a-strong representation of (2) can be computed using
 $$\exp(O(\sqrt{\log n\log \log n}))$$
multiplications.

\end{corollary}

The following result is the basis of our theorems in the present paper.

\begin{theorem}[\cite{G-mtxprod}]\label{dot}
\begin{itemize}
\item[(i)] Let $m=p_1p_2$, where $p_1\ne p_2$ are primes. Then a degree-2 1-a-strong representation of the dot-product
$$f(x_1,x_2,\ldots,x_n,y_1,y_2,\ldots,y_n)=\sum_{i=1}^n x_iy_i$$
can be computed with 
$$\exp(O(\sqrt{\log n\log \log n}))\eqno{(3)}$$
multiplications.

\item[(ii)] Let the prime decomposition of $m=p_1^{e_1}p_2^{e_2}\cdots p_r^{e_r}$. Then a degree-2 1-a-strong representation of the dot-product $f$ modulo $m$ can be computed using
$$\exp\bigg(O\bigg({\sqrt[r]{\log n(\log \log n)^{r-1}}}\bigg)\bigg)\eqno{(4)}$$
multiplications.

\item[(iii)] Moreover, the representations of (i) and (ii) can be computed on bilinear $\Sigma\Pi\Sigma$ circuits of size (3), and (4), respectively.
\end{itemize}
\end{theorem}

We reproduce here the short proof:

\begin{proof}
Let $g(x,y)=g(x_1,x_2,\ldots,x_n,y_1,y_2,\ldots,y_n)$ be the degree-2 polynomial from Theorem \ref{fo-1} which is a 0-a-strong representation of $S_n^2(x,y)$. Then consider polynomial
$$h(x,y)=(x_1+x_2+\ldots+x_n)(y_1+y_2+\ldots+y_n)-g(x,y)\eqno{(5*)}.$$
In $h(x,y)$, the coefficients of monomials $x_iy_i$ are all 1's modulo $m$, and the coefficients of monomials $x_iy_j$, for $i\ne j$ are 0 at least for one prime-power divisor of $m$. Consequently, by Definition \ref{a-strong}, $h(x,y)$ is a 1-a-strong representation of the dot-product $f(x,y)$.
\end{proof}

The following definition is a natural generalization of the a-strong representations for matrices:

\begin{definition}
Let $A=\{a_{ij}\}$ and $B=\{b_{ij}\}$ be two $n\times n$ matrices over $Z_m$. Then $C=\{c_{ij}\}$ is the alternative (1-a-strong, 0-a-strong) representation of the matrix $A$ if for $1\leq i,j \leq n$, the polynomial $c_{ij}$ of $n^2$ variables $\{a_{ij}\}$ is an alternative (1-a-strong, 0-a-strong) representation of polynomial $a_{ij}$. 

Consequently, we say that matrix $D=\{d_{ij}\}$ is an alternative (1-a-strong, 0-a-strong) representation of the product-matrix $AB$, if for $1\leq i,j \leq n$, $d_{ij}$ is an alternative (1-a-strong, 0-a-strong) representation of polynomial
$$\sum_{k=1}^n a_{ik}b_{kj}$$
modulo $m$, respectively.

\end{definition}

In \cite{G-mtxprod} we proved by $n^2$ applications of Theorem \ref{dot} that the 1-a-strong representations modulo 6 of the product of two $n\times n$ can be computed with $n^{2+o(1)}$ multiplications. Here we significantly improve this result: we show that the 1-a-strong representation can be computed by $n^{o(1)}$ multiplications. Moreover, the computation can be performed on a depth-3 homogeneous, bi-linear $\Sigma\Pi\Sigma$ circuit (bi-linear means that $\ell=2$ in   (1)). 

Let us stress that we emphasize that the computation can be performed on such a circuit because this circuit is perhaps the simplest model of computation in which matrix-product can be computed; the main result is that a certain representation of the matrix product can be computed by so few multiplications.

\section{Our result for matrix compression}

For simplicity, we prove our result here only for modulus 6; for other moduli, the poof is very similar, using Theorem \ref{dot}, part (ii) instead of part (i). 

If we do not say otherwise, the computations are modulo 6.

By Theorem \ref{dot}, a 1-a-strong representation of the dot-product 
$\sum_{i=1}^n x_iy_i$
can be computed  as 
$$\sum_{i=1}^n x_iy_i+3g(x,y)+4h(x,y)=\sum_{j=1}^t\left(\sum_{i=1}^nb_{ij}x_i\right)\left(\sum_{i=1}^nc_{ij}y_i\right)\eqno{(*1)}$$
where $b_{ij}, c_{ij}\in\{0,1\}$ and where both $g$ and $h$ has the following form:
$\sum_{i\ne j}a_{ij}x_iy_j,\ \ a_{ij}\bmod 6\in\{0,1\},$
and no term $x_iy_j$ appears in both $f$ and $g$; and $t=\exp(O(\sqrt{\log n\log \log n}))=n^{o(1)}$.

Now, let us observe that for each $j=1,2,\ldots,t$, 
$$\sum_{i=1}^nb_{ij}x_i\eqno{(*2)}$$
is a linear combination of variables $x_i$. If we plug in vectors instead of just variables in these homogeneous linear forms, then we will get linear combinations of the vectors.

Now, let $X$ denote an $n\times n$ matrix with entries as variables $\{x_{ij}\}$, and let $x_i$ denote its $i^{\tt th}$ column, for $i=1,2,\ldots,n$ and let $B=\{b_{ij}\}$, an $n\times t$ matrix of (*1) and let $C=\{c_{ij}\}$ an $n\times t$ matrix of (*1). 

The columns of the $n\times t$ matrix-product 
$$XB$$
gives the linear combinations specified by (*2). However,
$$XBC^T$$
is an $n\times n$ matrix, such that its column $\nu$ is equal to 
$$x_\nu+3g_\nu(X)+4h_\nu(X)$$
where $g_\nu(X)$ and $h_nu(X)$ are linear combinations of the columns of $X$ such that none of which contains $x_\nu$ and they do not contain the same column with non-zero coefficients. The proof of this fact is obvious from (*1), observing that
plugging in 
$$y^\nu=(0,0,\ldots,\overbrace{1}^\nu,0,\ldots,0)$$ we simply generate some linear combinations of the columns of matrix $XB$, and the coefficient in these combinations are nothing else that the rows of $C$.

Consequently, we proved the following implication of Theorem \ref{dot}:

\begin{theorem}\label{fo-1}
For any non-prime-power $m>1$ and for the $n\times n$ matrix $X=\{x_{ij}\}$, there 
exist effectively computable constant $n\times t$ matrices $B$ and $C$, such that 
$$XBC^T$$
is a 1-a-strong representation of matrix $X$ modulo $m$, where $t$ is equal to quantity (4), that is, $t=n^{o(1)}$. Moreover, every column of $XBC^T$ is a linear combination of the columns of $X$.
\end{theorem}

The dimension-defying implication of Theorem \ref{fo-1} is that $X$ is an $n\times n$ matrix, $XB$ is an $n\times n^{o(1)}$ matrix, and $XBC^T$ is again an $n\times n$ matrix.

An easy corollary of Theorem \ref{fo-1}, that 

\begin{corollary}\label{egy}
With the notations of Theorem \ref{fo-1},
$CB^TX$ is a 1-a-strong representation of matrix $X$ modulo $m$, where $t$ is equal to quantity (4), that is, $t=n^{o(1)}$. Moreover, every row of $CB^TX$ is a linear combination of the rows of $X$.
\end{corollary}

Our main result in this section is the following 

\begin{theorem}\label{fo-2}
For any non-prime-power $m>1$ and for the $n\times n$ matrix $X=\{x_{ij}\}$, there 
exist effectively computable constant $n\times t$ matrices $B$ and $C$, such that 
$$B^TXB$$
is a $t\times t$ matrix,  where $t$ is equal to quantity (4), that is, $t=n^{o(1)}$, and matrix
$$CB^TXBC^T$$
is a 1-a-strong representation of matrix $X$ modulo $m$.
\end{theorem}

The dimension-defying implication of Theorem \ref{fo-2} is that from the $n\times n$ matrix $X$ with simple linear transformations we make the tiny $n^{o(1)}\times n^{o(1)}$ matrix $B^TXB$, and from this, again with simple linear transformations, $n\times n$ matrix $CB^TXBC^T$, where it is a 1-a-strong representation of matrix $X$ modulo $m$.

\begin{proof}
From Theorem \ref{fo-1}, $XBC^T$ is a 1-a-strong representation of matrix $X$ modulo $m$. Moreover, every column of $XBC^T$ is a linear combination of the columns of $X$. By Corollary \ref{egy}, for any $n\times n$ $Y$, $CB^TY$ is a 1-a-strong representation of matrix $Y$ modulo $m$, and, every row of $CB^TY$ is a linear combination of the rows of $Y$. Plugging in $Y=XBC^T$, we get that the matrix $CB^TXBC^T$ is a 1-a-strong representation of the matrix $XBC^T$, moreover, its rows are linear combinations of the rows of $XBC^T$, that is, $CB^TXBC^T$ is also a 1-a-strong representation of the original $X$.\qed

\end{proof}

\section{Our result for matrix multiplication}

\subsection{Preliminaries}

We need to define a sort of generalization of matrix-product:

\begin{definition}\label{strange}
$f:R^{2n}\to R$  is a homogeneous bilinear function over ring $R$ if 
$$f(x_1,x_2,\ldots,x_n,y_1,y_2,\ldots,y_n)=\sum_{1\leq i, j\leq n}a_{ij}x_iy_j$$
for some $a_{i,j}\in R$. Let $U=\{u_{ij}\}$ be an $u\times n$ matrix over ring $R$, and let $V=\{v_{k\ell}\}$ be an $n\times v$ matrix over $R$. Then 
$$U(f)V$$
denotes the $u\times v$ matrix over $R$ with entries $w_{i\ell}$, where
$$w_{i\ell}=f(u_{i1},u_{i2},\ldots,u_{in},v_{1\ell},v_{2\ell},\ldots,v_{n\ell}).$$ 
\end{definition}

First we need a simple lemma, stating that the associativity of the matrix multiplication is satisfied also for the ``strange'' matrix-multiplication defined in Definition \ref{strange}:

\begin{lemma}\label{assoc}
Let $f(x_1,x_2,\ldots,x_n,y_1,y_2,\ldots,y_n)=\sum_{1\leq i, j\leq n}a_{ij}x_iy_j$ and $g(x_1,x_2,\ldots,x_v,y_1,y_2,\ldots,y_v)=\sum_{1\leq i, j\leq v}b_{ij}x_iy_j$ be homogeneous bilinear functions over the ring $R$. Let $U=\{u_{ij}\}$ be an $u\times n$ matrix, and let $V=\{v_{k\ell}\}$ be an $n\times v$ matrix, and $W=\{w_{ij}\}$ be a $v\times w$ matrix over $R$, where $u,n,w$ are positive integers.
Then
$$\left(U(f)V\right)(g)W=U(f)\left(V(g)W\right),$$
that is, the ``strange'' matrix-multiplication, given in Definition \ref{strange}, is associative. 
\end{lemma}

{\noindent \bf Proof 1:} The proof is obvious from the homogeneous bi-linearity of $f$ and $g$.

{\noindent \bf Proof 2:} We also give a more detailed proof for the lemma. The entry of row $i$ and column $k$ of matrix $U(f)V$ can be written as
$$\sum_{z,t}a_{zt}u_{iz}v_{tk}.$$
Consequently, the entry in row $i$ and column $r$ of $(U(f)V)(g)W$ is
$$\sum_{k,\ell}b_{k\ell}\left(\sum_{z,t}a_{zt}u_{iz}v_{tk}\right)w_{\ell r}.$$

On the other hand, entry $(t,r)$ in $V(g)W$ is
$$\sum_{k,\ell}b_{k\ell}v_{tk}w_{\ell r},$$
and entry $(i,r)$ in $U(f)(V(g)W)$ is
$$\sum_{z,t}a_{zt}u_{iz}\sum_{k,\ell}b_{k\ell}v_{tk}w_{\ell r},$$
and this proves our statement.\qed

Now we are in the position of stating and proving our main theorem for matrix multiplications:

\begin{theorem}\label{main}
Let $X$ and $Y$ two $n\times n$ matrices, and let $m>1$ be a non-prime-power integer. Then the 1-a-strong representation of the matrix-product $XY$ can be computed with $t=n^{o(1)}$ multiplication, where $t$ is given by (4).
\end{theorem}

\begin{proof}
We use Theorem \ref{fo-1} and Corollary \ref{egy}. Let us consider 
$t\times n$ matrix $B^TX$ and $t\times n$ matrix $YB$; these matrices can be computed without any multiplications from $X$ and $Y$ (we do not count multiplications by constants). Let $h(x,y)$ be the homogeneous bi-linear function (5*). Then 
$$B^TX(h)YB$$
can be computed with $n^{o(1)}$ multiplications (Note, that because of Lemma \ref{assoc}, the associativity holds). 
Now compute matrix
$$CB^TX(f)YBC^T=(CB^TY)(f)(YBC^T)$$
without any further (non-constant) multiplication. By Theorem \ref{fo-1} and Corollary \ref{egy}, $CB^TX$ and $YBC^T$ is a 1-a-strong representations of $X$ and $Y$ respectively, and they are the linear combinations of the rows of $X$ and columns of $Y$, respectively. Consequently, using Theorem \ref{dot}, $CB^TX(f)YBC^T$ is a 1-a-strong representation of $XY$.\qed
\end{proof}

\noindent {\bf Acknowledgment.}

  The author acknowledges the partial support of the Széchenyi István fellowship.

%\bibliography{c:/vince/cikkek/gv}
%\bibliographystyle{c:/programs/pctex/texinput/alpha}

\end{document}